# Kernel Architecture of the Genetic Circuitry of the *Arabidopsis* Circadian System

Mathias Foo[1,2], David E. Somers[3], Pan-Jun Kim[1,4]*

[1]Asia Pacific Center for Theoretical Physics, Pohang, Gyeongbuk, Republic of Korea

[2]School of Engineering, University of Warwick, Coventry, United Kingdom

[3]Department of Molecular Genetics, The Ohio State University, Columbus, Ohio, United States of America

[4]Department of Physics, Pohang University of Science and Technology, Pohang, Gyeongbuk, Republic of Korea

*Corresponding author.
E-mail: pjkim@apctp.org

## Abstract

A wide range of organisms features molecular machines, circadian clocks, which generate endogenous oscillations with ~24 h periodicity and thereby synchronize biological processes to diurnal environmental fluctuations. Recently, it has become clear that plants harbor more complex gene regulatory circuits within the core circadian clocks than other organisms, inspiring a fundamental question: are all these regulatory interactions between clock genes equally crucial for the establishment and maintenance of circadian rhythms? Our mechanistic simulation for *Arabidopsis thaliana* demonstrates that at least half of the total regulatory interactions must be present to express the circadian molecular profiles observed in wild-type plants. A set of those essential interactions is called herein a *kernel* of the circadian system. The kernel structure unbiasedly reveals four interlocked negative feedback loops contributing to circadian rhythms, and three feedback loops among them drive the autonomous oscillation itself. Strikingly, the kernel structure, as well as the whole clock circuitry, is overwhelmingly composed of inhibitory, rather than activating, interactions between genes. We found that this tendency underlies plant circadian molecular profiles which often exhibit sharply-shaped, *cuspidate* waveforms. Through the generation of these cuspidate profiles, inhibitory interactions may facilitate the global coordination of temporally-distant clock events that are markedly peaked at very specific times of day. Our systematic approach resulting in experimentally-testable predictions provides insights into a design principle of biological clockwork, with implications for synthetic biology.

## Author Summary

Sleep/wake cycles in animals exemplify daily biological rhythms driven by internal molecular clocks, circadian clocks, which are important for plant life as well. The plant circadian clock is highly complex, eluding our understanding of its design principle. Based on the computational simulation of *Arabidopsis thaliana*, we successfully identified a *kernel* of the plant circadian system, the critical genetic circuitry for clock function. The kernel integrates four major negative feedback loops that process molecular circadian oscillations. Surprisingly, the plant clock circuitry was found to be overwhelmingly composed of inhibitory, rather than activating, interactions among genes. This fact underlies plant circadian molecular profiles to often exhibit sharply-shaped, *cuspidate* waveforms, which indicate clock events that are markedly peaked at very specific times of day. Our work presents experimentally-testable predictions, with implications for synthetic biology.

## Introduction

A variety of living organisms on Earth features built-in molecular clock machineries that control the organism's daily activities [1]. These internal time-keepers, circadian clocks, generate endogenous oscillations of gene expression with ~24 h periodicity, enabling the anticipation of diurnal environmental variations and the coordination of biological processes to the optimal times of day. Examples of such biological processes include sleep/wake cycles in animals, emergence from the pupal case in fruit flies, spore formation in fungi, and leaf movements in plants [2-4]. Disruption of circadian rhythmicity is associated with a wide range of pathophysiological conditions, indicating the importance of clock functions in homeostasis [5-8]

Compared to other organisms, such as fungi, insects, and mammals whose circadian systems have been well studied, a molecular understanding of the plant circadian system is still elusive. Numerous molecular and genetic approaches using *Arabidopsis thaliana* have facilitated the discovery of more than 20 plant clock genes as well as their regulatory interactions [1, 9, 10]. The emerging picture from this effort suggests that the core regulatory circuit of the plant circadian system is more complex than in other organisms [9, 11-13]. The apparent complexity of the plant clock machinery raises a fundamental question: are all the regulatory interactions between clock genes equally necessary for the establishment and maintenance of plant circadian rhythms? In other words, can we distinguish more important from less important regulatory interactions for normal clock functioning? Answering this question involves an attempt to prioritize our focus amongst numerous regulatory interactions, in order to simplify a global view of, and thereby elicit an essential principle of, the plant clock organization. Despite the fundamental importance of this issue, a satisfactorily systematic approach has not been taken yet; thus, this topic is the focus of our study. In the case of other biological processes, finding essential subnetworks out of the whole has been of wide interest for both scientific and engineering purposes [14-18].

Properly designed experiments may be one way to address this issue, but often require laborious and costly efforts. Complementary to experiments, mathematical models help biological findings by predicting the effects of genetic and non-genetic perturbations, where experimental access could be limited or unavailable. Utility of mathematical models has been well documented in earlier studies of circadian rhythms [19-22]. An initial mathematical model of the plant circadian system was constructed based only on three genes, *LATE ELONGATED*

*HYPOCOTYL* (*LHY*), *CIRCADIAN CLOCK ASSOCIATED 1* (*CCA1*), and *TIMING OF CAB EXPRESSION 1* (*TOC1*) [22]. This model has evolved to include five times more components to date [23, 24]. Additionally, models that incorporate the downstream targets of the core circadian system are starting to gain attention [25]. These models have certainly served a significant role in enhancing our understanding of the plant circadian clock. Nevertheless, to the best of our knowledge, none of these studies has fully attempted to specify the functionally essential interactions between clock genes in a systematic and comprehensive way.

Central to our approach to the plant circadian system is the concept of a *kernel*. We define a kernel as a collection of minimal functional sets, each comprising all molecular components (genes and gene products) in the system and only a part of their regulatory interactions, which must be present to generate the temporal trajectory of molecular concentrations close to wild type (WT). In this definition, we refer to a collection of minimal sets to cover cases with multiple minimal sets. Based on an *Arabidopsis* clock model constructed in this study, our analysis shows that the kernel structure combines four negative feedback loops whose interplay effectively accounts for circadian rhythmicity in *Arabidopsis*. Strikingly, the kernel structure, as well as the whole clock circuitry, was found to be overwhelmingly composed of inhibitory interactions between genes. We subsequently present a mechanistic reason for the prevalence of such inhibitory interactions in the plant clock. These results provide a systematic and unique view of the plant circadian oscillators, with experimentally testable predictions to enhance our understanding of biological time.

## Results

**Construction and verification of the mathematical plant clock model**

We began by constructing a mathematical model of the core circadian oscillator in plant *Arabidopsis thaliana*. For this model construction, we applied system identification techniques to publicly available time course data of mRNA and protein expression (Materials and Methods). The resulting model consists of 24 ordinary differential equations (ODEs), describing a rate of a concentration change of each mRNA, protein, or protein complex (S1 Text). Experimentally-verified molecular interactions were primarily incorporated in the model, which then contains a total of 40 transcriptional and post-translational interactions between components, along with light-dependent regulations. Fig 1A shows a global architecture of the core gene circuit considered in our model.

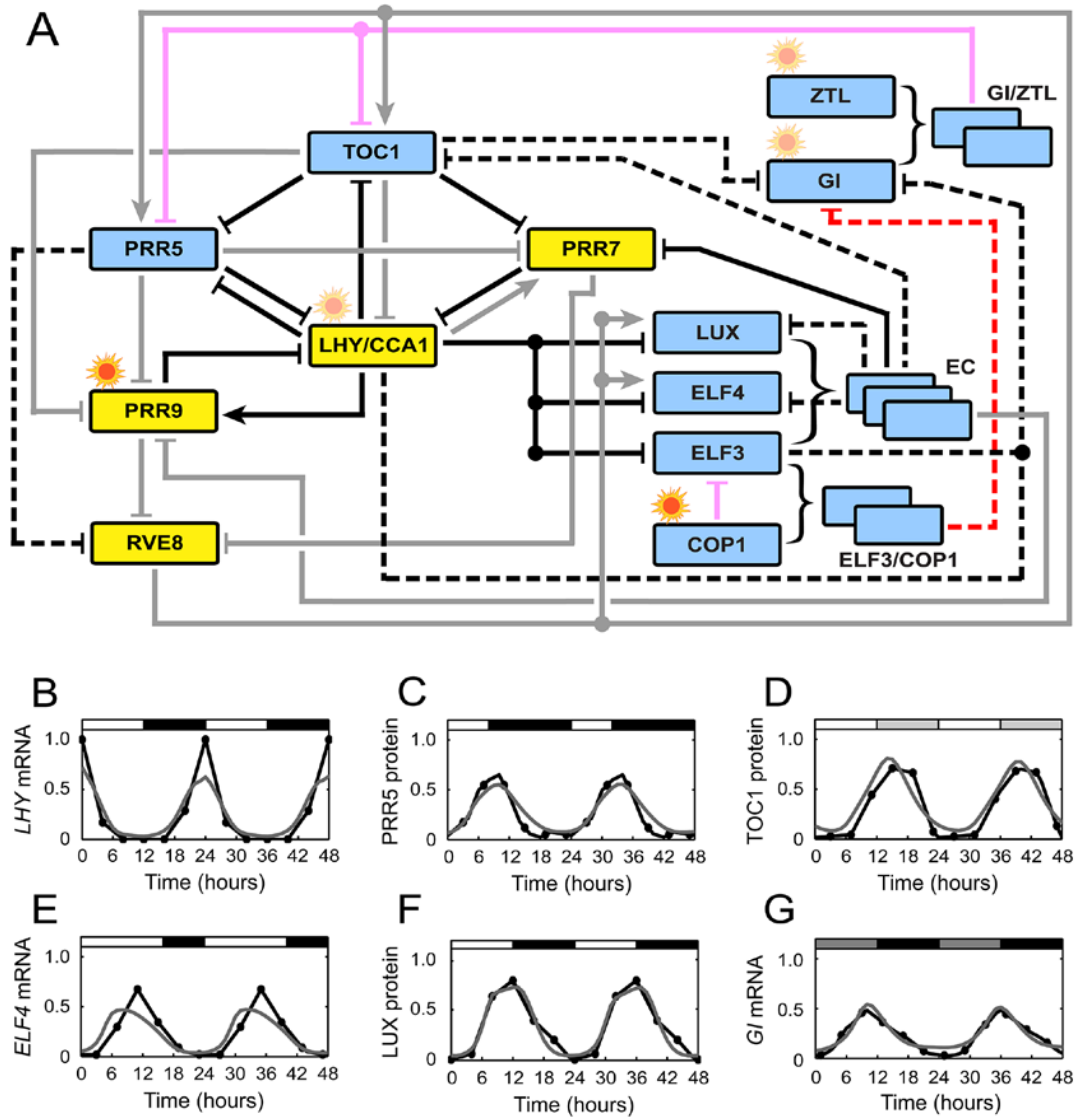

**Fig 1. Core of the *Arabidopsis* circadian clock.** (A) Regulatory circuits of clock components in the MF2015 model. Boxes denote molecular components (yellow and blue for morning and evening components, respectively). Lines denote activating (arrow-headed) or inhibitory (bar-headed) regulation, whether transcriptional or post-translational (described below in detail). Light-dependent regulation is denoted by a sun-like symbol on each box. Curly brackets indicate the formation of protein complexes. Black and red lines constitute the kernel structure and represent transcriptional (black) and post-translational (red) regulation; among them, solid lines belong to the four major negative feedback loops,

whereas dashed lines do not belong to those loops. Gray and pink lines do not constitute the kernel structure, and represent transcriptional (gray) and post-translational (pink) regulation. Among light-dependent regulations (sun-like symbols), only those of *PRR9* and COP1 (sun-like symbols with stronger colors) belong to the current kernel structure. (B–G) Comparison between experimental (black) and simulated (gray) mRNA or protein levels under different light conditions. (B, F) 12L:12D cycles. (C) Short days. (D) LL. (E) Long days. (G) DD. In (B, C, E, F), white and black segments correspond to light and dark intervals, respectively. In (D, G), relatively light and dark segments indicate subjective days and nights, respectively. The sources of the experimental data in (B–G) are presented in S1 Table.

In comparison with previous models [23, 24, 26], the new model is mainly based on the model (P2013) by Pokhilko et al. [23], but we filtered out hypothetical or outdated molecular interactions and adopted some recent findings [24]. Compared to our earlier work [26], which uses a discrete-time model for control design purposes, here we have constructed a continuous-time model, with revised interactions compatible with recent knowledge. Full details of the model comparisons are presented in S1 Text. Overall, we stress that our current model does not intend to outperform other existing models in its accuracy through the inclusion of all up-to-date information. Rather, the priority was to construct a model which is compact, yet biologically relevant, in accordance with recent experimental knowledge. We expect that this model is suitable enough for our main purpose of kernel identification, without further sophistication of the model structure.

Because we are ultimately moving forward to identify the kernel structure responsible for circadian rhythms in WT plants, time series data of mRNA and protein expression from WT, not from mutants, were used during model construction. Mutant data were used only to validate the constructed model, as will be described later. Specifically, we estimated the parameters of the model by fitting the simulation results to WT mRNA and protein expression profiles over time, under five different light conditions: equal length light-dark cycle, i.e., 12 hours of light and 12 hours of dark (12L:12D), 16 hours of light and 8 hours of dark (long day), 8 hours of light and 16 hours of dark (short day), constant light (LL), and constant dark (DD). These expression profiles were obtained from publicly available experimental literature and databases (Table S1). Because the absolute levels of mRNAs and proteins were difficult to ascertain from their sources, we normalized the expression levels into dimensionless values (≤1) with arbitrary scales. As a proxy for the *LHY/CCA1* information, we adopted the *LHY* expression data, because they were often better in the quality than *CCA1*'s. Constraining the model output to fit all these datasets gave rise to a total of 97 estimated parameters of the model equations, along with 51 coefficients that scale each light condition's mRNA and protein levels relative to the levels under 12L:12D cycles (see S1 Text). Our model does not separate nuclear from cytosolic proteins [27, 28], due to incomplete availability of the relevant expression data and to avoid increasing model complexity.

What is the resulting performance of our model (MF2015)? We found that MF2015 captures well the overall temporal patterns of gene expression from WT (Fig 1B–1G; for comparison with P2013, see S1 Text). Also, the free running rhythms in WT are in good agreement with experimental values [29, 30]: 25.2 h (model) and 24.6 h (experiment) in LL, and 25.8 h (model) and 25.9 h (experiment) in DD. However, these results cannot validate MF2015, because we estimated the model parameters from the WT data. To directly test the predictive power of the model against an independent dataset, we computed the altered rhythmicity under different genetic perturbations. The simulated mutants are 76.2% accurate when the clock periods are quantitatively compared to experimental values (see S1 Text). Qualitative agreement (lengthened period, shortened period, or arrhythmia) is observed for 85.7% of the simulation outcomes and experimental results (S1 Text). Moreover, the simulation predicts the substantial elevation (reduction) of ZEITLUPE (ZTL) protein levels in LL (DD), matching the experimental finding [31]. This result is the first accurate reproduction of ZTL performance through computational modeling (S6 Fig). Taken together, MF2015 is greatly supported by an array of experimental evidence in terms of its predictability. Note that P2013 yields the simulated mutant periods in 42.9% quantitative agreement with experimental values.

In general, the simulation outcomes were robust to a wide range of kinetic parameter variations and transient molecular concentration changes (S1 Text). A few exceptions that convey the system's sensitive response involve the variations of parameters in *PSEUDO RESPONSE REGULATOR 5* (*PRR5*) mRNA degradation, *EARLY FLOWERING 3* (*ELF3*) inhibition by LHY/CCA1, and light-responsive protein production. Whether they represent genuine biological factors or model incompleteness is unknown. Meanwhile, the overall robustness to parameter variations indicates the presence of multiple parameter sets for the model. Interestingly, alternative parameters that we examined did not make much of an improvement in the predictability of mutant period lengths (S1 Text). Moreover, such alternative parameters of the model are unlikely to change the main results of our study, as kernel identification and analysis involve parameter re-optimization processes.

**Kernel identification from the plant circadian system**

Our modeling of the core circadian system (MF2015) encouraged us to address difficult mechanistic questions. Among all 40 molecular interactions and light regulations in the system, which interactions (and light regulations) are minimally necessary to shape the circadian mRNA and protein expression profiles observed in WT across different light conditions? We refer to this collection of minimal sets as the kernel of the circadian system. In the next paragraph, both molecular interactions and light regulations are referred to simply as interactions.

Sheer screening of interaction sets, whose removal severely distorts clock rhythmicity, would not be sufficient to identify a kernel structure. If this distortion is repaired by a readjustment of kinetic parameters, the removed interactions are not likely to be essential in their network-topological properties; rather, their knockout effect is simply dependent on specific parameters. Therefore, the knockout effect in distorting clock rhythms should be double-checked with re-optimized parameters. If the knockout effect remains severe even after parameter re-optimization, the removed interactions can now be said to be essential in their network topological properties. Ideally, our kernel discovery procedure would be to search through all possible combinations of interactions, and examine the effects when the interactions in each combination are removed, followed by parameter re-optimization to best fit the WT expression profile of every clock component across different light conditions. This strategy, although ideal, is extremely computationally demanding and therefore impractical. Instead, we devised a heuristic approach that consists of the following steps (Materials and Methods, and S1 Text): first, we measure the knockout effect of each interaction on the WT expression patterns under the five different light conditions. Then, we prune those interactions from weak to strong knockout effects until discovering any single clock component that fails to produce rhythms similar to WT. Next, among the remaining interactions, we choose those with knockout effects below a certain threshold. Each chosen interaction is deleted, and parameter re-optimization follows to fit the WT expression data. If parameter re-optimization recovers the WT rhythms for every clock component, this interaction is completely removed from the system. The implementation of these steps, complemented by an additional step to allow multiple solutions, leaves a fraction of the interactions, which yet connect all the molecular components in the system. This interaction set corresponds to our estimated kernel structure. For a detailed description of the kernel identification, see S1 Text.

Using MF2015, we found that the kernel of the plant circadian system consists of 22 transcriptional and post-translational interactions and light regulations, which seamlessly involve all molecular clock components in the system. In other words, at least half of the 40 interactions/regulations in the whole system are required to form the WT rhythms across the five different light conditions. Notably, the kernel structure harbors four negative feedback loops, termed loops I to IV (Fig 2; compare with Fig 1A). In the kernel, the only negative feedback other than these four loops is the autoinhibition of the EVENING COMPLEX (EC) genes through the EC, and this effect remains localized to the EC formation and thus not our focus here. Loops I to IV host at least one of the *PSEUDO RESPONSE REGULATOR* (*PRR*) genes each, and are interlocked by having *LHY/CCA1* in common: loop I includes *LHY/CCA1*, *PRR5*, and *TOC1* (Fig 2A). Loop II has *LHY/CCA1*, *PRR7*, and *TOC1* (Fig 2B). Loop III involves *LHY/CCA1*, *PRR7*, and the EC, along with the EC subcomponents (Fig 2C). Lastly, loop IV includes *LHY/CCA1*, and *PRR9* regulated by light (Fig 2D). Accordingly, *TOC1* interconnects loops I and II, while *PRR7* interconnects loops II and III. Each of loops I, II, and III includes a cyclic structure of triple inhibitions, known as a repressilator (Fig 2A–2C) [32]. A repressilator structure can exhibit sustained oscillation under proper conditions. Of note, loop I has one more interaction added to this repressilator structure, i.e., the inhibition of *PRR5* by LHY/CCA1. The direction of this inhibitory interaction is exactly opposite to the repressilator's overall cyclic direction, and thus is supposed to be antagonistic to the oscillatory capability of the loop (see below). Among the four loops, loop IV in Fig 2D is the simplest one, having only a pair of single positive and negative connections between two morning-expressed components, coupled with light.

To our knowledge, loops I and II have not been previously described, whereas loop III recapitulates a repressilator structure previously reported [33]. Loop IV has been previously termed the morning loop [9, 34, 35]. Therefore, our unbiased and systematic approach to kernel identification does not only recover previously characterized gene circuits (loops III and IV), but also suggests new circuits (loops I and II) that may be crucial for *Arabidopsis* clock function.

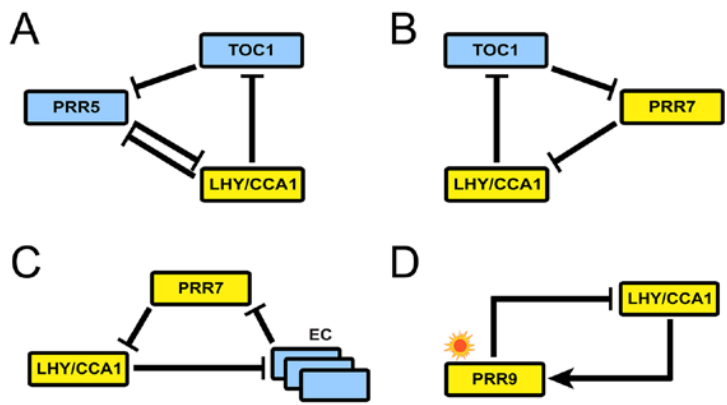

**Fig 2. Four major negative feedback loops in the kernel.** (A) Loop I. (B) Loop II. (C) Loop III. For visual clarity, the detailed EC subcomponents are omitted (available in Fig 1A). (D) Loop IV. In (A–D), the same symbols were used as in Fig 1A.

**Dynamical capability of the kernel-embedded feedback loops**

Owing to the above kernel identification, the complex plant clock circuitry has been greatly simplified, converging on the four negative feedback loops that structure the kernel. We next considered an in-depth mechanistic analysis of the individual feedback loops as well as their interrelations.

An immediate question is, among the four negative feedback loops, which of the loops critically support the generation of *autonomous* molecular oscillations observed in WT. By definition, every element in the kernel must play a significant role in shaping the oscillatory profiles. However, it does not mean that their contributions to the creation of the autonomous oscillation are necessarily equivalent to each other. Moreover, the current kernel structure is a full repertoire of interactions necessary for all five different light conditions mentioned above. Clearly, only separate simulations of constant, free running conditions will answer this question for the endogenous, autonomous oscillation.

To test the capability of individual loops to generate autonomous oscillations close to WT, we simulated LL using a computational model of each isolated loop, with kinetic parameters re-optimized for the WT expression data in LL (S1 Text). Given the WT expression profiles, this parameter re-optimization was expected to reveal the maximum oscillatory capacity of each loop structure regardless of its specific MF2015 parameters. It infers a natural bound of the loop's contribution to the WT endogenous oscillations − a natural bound imposed by the loop's structure itself rather than by specific parameters. From this simulation, we found that loops I, II, and III in LL were clearly able to generate sustained oscillations similar to WT (Fig 3A and 3B), whereas loop IV failed (Fig 3C). In fact, if equipped with other parameters, oscillations can be maintained even by loop IV, but at the expense of its specific oscillatory patterns, in far deviation from the experimental profiles. Once loop IV undergoes a parameter adjustment to fit the experimental profiles, it loses sustained oscillation.

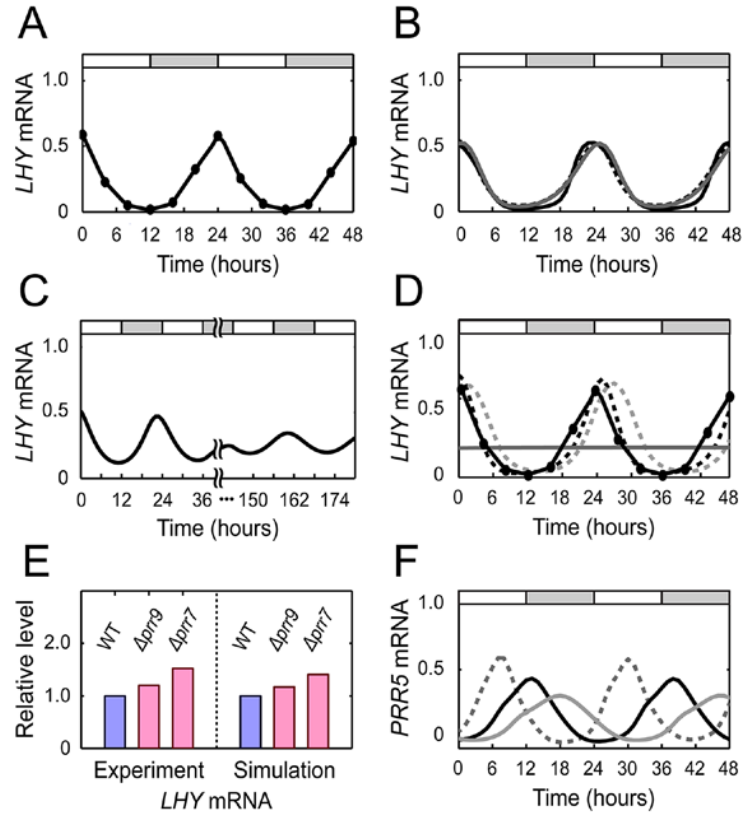

**Fig 3. Dynamical properties of the kernel.** (A) Experimental *LHY* mRNA levels in LL (Table S1). (B, C) *LHY* mRNA levels from each simulation of loops I to III (B) and loop IV (C) in LL. The parameters of each loop were re-optimized for WT expression data in LL. Black solid, black dotted, and gray solid lines in (B) are for loops I, II, and III, respectively. (D) The same *LHY* mRNA levels in (A), along with simulated *LHY* mRNA levels when *LHY/CCA1* inhibition by PRR5 (black dotted), or by PRR7 (gray dotted), or by both PRR5 and PRR7 (gray solid) was removed from MF2015. The MF2015 simulation was performed in LL with re-optimized parameters. (E) *LHY* mRNA levels from the *Δprr9* and *Δprr7* knockout mutants in LL (time averages normalized with respect to WT data). Experimental values [29] and MF2015 simulation results are compared. (F) MF2015-simulated *PRR5* mRNA levels in LL when *PRR5* inhibition by LHY/CCA1 is increased (gray solid) or decreased (gray dotted) by 20%, or is not adjusted (black solid). In (A–D, F), white and gray segments indicate subjective days and nights, respectively.

---

The endogenous oscillatory capability of individual loops I to III raises an intriguing possibility: can the plant circadian rhythm be robust to the breakage of some loop(s), if buffered by the other loop(s)' activity? To explicitly address this question, we inactivated loop I in MF2015 by blocking the inhibition of *LHY/CCA1* by PRR5. Likewise, we inactivated both loops II and III simultaneously, by blocking the inhibition of *LHY/CCA1* by PRR7. The MF2015 simulation of LL demonstrates that either of these two "mutations" largely restores the circadian gene expression profiles observed in WT, if accompanied by parameter re-optimization (Fig 3D). As can be predicted, the simultaneous blockage of both PRR5 and PRR7's inhibitory actions on *LHY/CCA1* in MF2015 inactivated all three oscillatory loops I to III, and thus abolished the circadian rhythmicity itself of gene expression, even when accompanied by parameter re-optimization. This prediction is well supported by an experimental report that the *Δprr5/prr7* double mutant in constant conditions exhibits almost arrhythmic mRNA levels of clock-controlled genes, although each single mutant retains free running rhythmicity [36]. Moreover, the above simulation forecasts that only the removal of the two inhibitory interactions, rather than the entire double gene deletion, is necessary to cause severely abnormal clock gene expression. In sum, we find that under certain circumstances loop I can buffer the loss of loops II and III, and vice versa. Similarly, we computationally blocked *PRR7* inhibition by TOC1, and that by the EC, to inactivate loop II and loop III, respectively. Again, simulated mutant outcomes suggest that loop II and loop III can buffer the loss of each other. Taken together, these results indicate complementary relationships between loops I, II, and III in the management of endogenous circadian oscillations.

While loops I to III exhibit the fundamental capacity to generate endogenous oscillations similar to WT, loop IV lacks such capability. We therefore conjectured that, among all the four loops, loop IV is unlikely to exert the strongest regulation on the clock gene expression, if these genes are regulated by the other loops as well. Indeed, the *LHY/CCA1* inhibition by PRR9 (in loop IV) was consistently weaker than either the *LHY/CCA1* inhibition by PRR7 or that by PRR5 (in loops I to III), throughout our simulation with various re-optimized parameters (S1 Text). Previous experimental data from LL have shown that a *Δprr9* knockout has a smaller effect on *LHY* and *CCA1* expression than a *Δprr7* knockout [29]. Fig 3E shows that *LHY* mRNA levels, on average, increased by 60.5% and 16.7% in the *Δprr7* and *Δprr9* mutants, respectively, consistent with our computational prediction; a similar trend was also observed for *CCA1* mRNA [29]. Despite the loop IV's relatively weak role in free running rhythmicity, it should be noted that, in our current kernel structure, loop IV is the only negative feedback loop which senses external light stimulus (Fig 2D) and thereby contributes to the entrainment of the kernel dynamics to light. We cannot entirely exclude the possibility that more loops may come into play in light sensing of the kernel as our model becomes updated.

The efficacy of our simple kernel structure to interpret the clock dynamics is further exemplified by loop I. In addition to the basic repressilator structure, loop I holds a unique topological feature of reciprocal inhibitory interactions between *LHY/CCA1* and *PRR5* (Fig 2A). In particular, the inhibition of *PRR5* by LHY/CCA1 is placed in opposition to the repressilator's overall cyclic direction, and thus may retard the loop's inherent oscillation. In fact, this retardation effect was found to affect the oscillation of the whole clock circuitry, because of the structural interconnection between loop I and the whole. For example, the simulation of MF2015 in LL demonstrates that a 20% increase in *PRR5* inhibition by LHY/CCA1 slows down the circadian rhythm, resulting in a 3.3 h lengthened period, whereas a 20% decrease in this inhibition shortens a period by 2.9 h (Fig 3F). This experimentally-testable idea might be hard to conceive without the simplicity of the loop-I structure.

In the kernel, *LHY/CCA1* interlocks all loops I to IV, indicating its central role in the circadian oscillator. The adverse effect of the *Δlhy/cca1* double knockout on model performance is supported by experimental evidence [37, 38]. From the entire kernel structure in Fig 1A, compared with loops I to III, one can notice the presence of *TOC1* inhibition by the EC. This inhibition is the only regulatory interaction with its regulated target (*TOC1*) in the loops, while the interaction itself is not a part of major negative feedback loops in the kernel. This fact prompted us to investigate whether *TOC1* inhibition by the EC should be retained in our kernel. The simulated removal of this inhibition from the kernel at least distorted the *LHY* mRNA and TOC1 protein profiles, even when accompanied by parameter re-optimization (S7 Fig). Therefore, we keep in the present kernel structure *TOC1* inhibition by the EC.

In conclusion, our model is supported by current experimental data and indicates that the plant circadian oscillator is an orchestrated interaction of mainly four negative feedback loops in the kernel. In the face of the larger complexity of the full circuitry, our simplified loop structures may offer an efficient way to understand the plant clock mechanisms, as well as predict circadian dynamics that has not yet been characterized.

**Prevalent inhibitory interactions and their functional advantage**

Among the four major negative feedback loops in the kernel, loops I to III have the repressilator-like structures that are entirely composed of inhibitory interactions. Only loop IV includes an activating interaction. Regarding the central role of these feedback loops in circadian rhythms, why does the plant circadian system favor such inhibitor-enriched loops for its function? Indeed, recent molecular studies of the plant circadian system have indicated that inhibitory relationships outnumber activating regulations among all clock genes [39]. The full circuitry considered in MF2015 is dominated by inhibitory interactions, and this feature becomes even more prominent in its kernel structure, harboring only one activating interaction (Fig 1A). The dominance of such inhibitory interactions distinguishes the plant clock from other circadian systems, including those of mammals and fungi, which have comparable numbers of inhibitory and activating interactions [11-13].

This issue can begin to be addressed by considering that the kernel structure is designed for the production of temporal gene expression patterns close to WT (Fig 4A). Therefore, we presumed that many inhibitory regulations, at least in the kernel, may generate specific waveforms of the WT expression profiles. We do observe, in fact, that a number of *Arabidopsis* clock genes often exhibit particular waveforms of mRNA and protein expression (Fig 1B–1G and S1–S5 Fig). This waveform is characterized by an asymmetry between the acrophase and bathyphase, as schematized in Fig 4B: the acrophase shows a relatively sharpened peak, whereas the bathyphase can be approximated as flat. Regarding the overall acuteness around a particular peak phase, we here describe this pattern as *cuspidate*. For comparison, a common sinusoidal wave is not cuspidate, having a symmetrically rounded shape to the acrophase and bathyphase.

To examine the possible relevance of inhibitory regulation in cuspidate waveforms, we created a mathematical system consisting of a single transcription factor, either an inhibitor or activator, and its own target gene (Fig 4A and Materials and Methods). We formulated the model equations similar to MF2015. On the assumption that the target gene shows a near cuspidate ~24h-period expression pattern of proteins (Fig 4B), we conversely asked what specific abundance profile the transcription factor (inhibitor or activator) should have for the production of that target gene profile. Our simulation results highlight a clear difference between inhibitor and activator cases, when the target gene exhibits a cuspidate pattern (S1 Text). The inhibitor or activator tends to have a large or small phase difference, respectively, of ~8 to 12 hours or ≲4 hours with the target gene in their protein profiles, as shown in Fig 4C–4E and S8 Fig. In other words, an inhibitor (activator) and its target have a roughly antiphase-like (inphase-like) relationship. Otherwise, the target gene's protein expression waveform will not be cuspidate but will exhibit a more smoothened profile (S9 Fig). These facts were initially observed in our simulation with simplified, yet realistic, protein expression profiles, such as that in Fig 4B. Even without such simplification, adopting empirical protein expression patterns for our simulation consistently supported the above results (S10 Fig). We also note that the cuspidate waveforms in the plant clock do not simply result from the sampling intervals of experimental data, as different interpolation methods for these data points (and the absence of such interpolation itself) gave similar profiles.

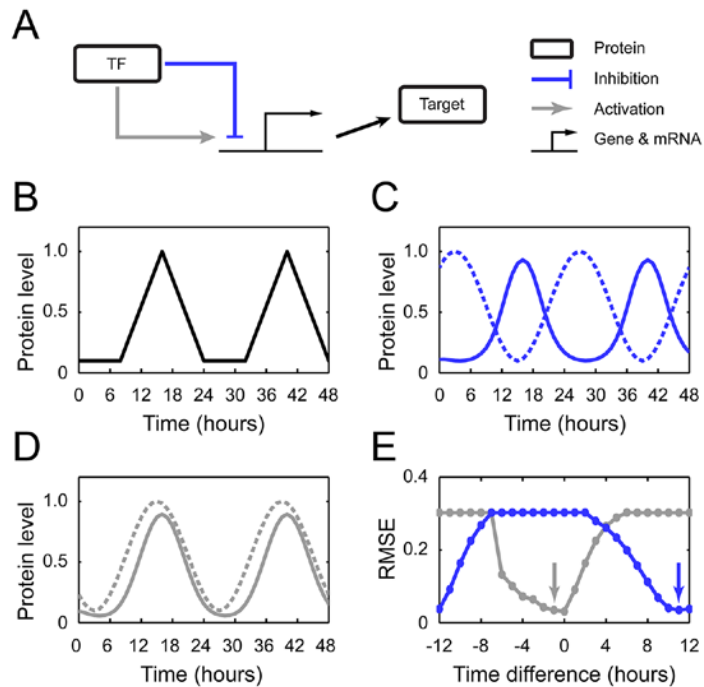

**Fig 4. Effect of an inhibitor or activator on the generation of cuspidate profiles.** (A) A transcription factor (TF; inhibitor or activator) regulates gene expression and thereby affects the protein production. (B) Schematic diagram of a cuspidate waveform. (C, D) Sinusoidal profile (dotted) of an inhibitor (C) or activator (D), which regulates the target gene to produce its proteins (solid). Given the inhibitor or activator levels, the target gene transcription, translation, and product degradation were simulated with the parameters that best fit the desired, cuspidate expression profile in (B). (E) For a given phase difference between a transcription factor (inhibitor or activator) and its target gene's protein, plotted is the resulting deviation of the target gene's protein profile from the cuspidate profile. The horizontal axis represents a peak time difference between each transcription factor's profile and the target gene's desired protein profile in (B): a sign is negative if the former profile has more advanced peak time than the latter, otherwise it is positive. The vertical axis represents a root mean square error (RMSE) between the target gene's actual and desired protein profiles when the target gene expression was simulated with the parameters that best fit the cuspidate protein profile in (B). The smaller the RMSE, the more cuspidate profile the target gene has. The inhibitor case is shown in blue, and the activator case is shown in gray. Arrows in blue and gray correspond to the conditions for (C) and (D), respectively. Sinusoidal waves were used for the transcription factor profiles as illustrated in (C, D). Qualitatively similar results are reproduced by other various waveforms (S8 Fig). For an alternative definition of peak time differences along the horizontal axis, use of the peak times of the target gene's actual, rather than desired, profiles did not essentially change the peak time differences.

---

Provided that a cuspidate profile confers accurate timing of biological events around the peak phase, what is the implication of our simulation results involving the cuspidate waveform and inhibitory or activating regulation? Inhibition-induced large phase differences between the genes correspond to the global coordination of multiple clock events, distant from each other in their peak times. Conversely, activation-induced small phase differences between the genes may coordinate only the clock events nearby in time. It is possible that activating regulation might also induce larger phase differences between the genes, but would not generate cuspidate profiles in this case (S9 Fig). This fact explains why the kernel does not keep the activating regulations by REVEILLE 8 (RVE8), whose target genes have large phase differences with RVE8, yet exhibit cuspidate profiles (hence, those profiles are presumably more attributed to other regulators of these target genes). To summarize, inhibitory interactions in the plant clock seem to support the temporal coordination of distant clock events peaked at very specific times. However, it should be stressed that inhibitory interactions do not necessarily result in cuspidate waveforms in all cases. Rather, obtaining such waveform profiles requires inhibitory interactions when involving genes with large phase differences in their peak expression. Employing the terms in propositional logic, the presence of both cuspidate waveforms and large phase differences is close to a sufficient condition to implicate inhibitory regulation as their cause, but is not the necessary condition. We also note that our current definition of a cuspidate waveform is largely qualitative, based on a particular type of asymmetry between the acrophase and bathyphase. Mathematically more rigorous characterization, along with the inclusion of other possible waveforms in our framework, deserves investigation.

Within the MF2015 kernel structure, a cuspidate-waveform gene which has multiple inhibitors tends to have larger phase differences with its strongest inhibitor, consistent with our framework. For example, the transcription of a cuspidate-waveform gene, *PRR5*, is repressed by both LHY/CCA1 and TOC1. There is a large phase difference between PRR5 and LHY/CCA1 proteins, ~8 h compared to the ~4 h difference between PRR5 and TOC1 proteins in a 12L:12D cycle. Supportively, MF2015 suggests that LHY/CCA1 inhibits *PRR5* expression ~17 times more than TOC1 (S1 Text). This fact indicates that the primary role of the *PRR5* inhibition by LHY/CCA1 is to ensure the PRR5's cuspidate waveform. Lowering the relative contribution of this inhibition (i.e., alleviating the repression by LHY/CCA1 while strengthening that by TOC1) reduces the peak-to-trough change in the PRR5 expression over time (performed under 12L:12D cycles to control for the periods of different expression profiles; see S11 Fig). Our analysis accounts well for why *PRR5* inhibition by LHY/CCA1 is present in the clock, although it is antagonistic to the system's overall oscillatory capability as noted previously in relation to loop I. However, we recognize that it may be hard to treat separately multiple transcription factors regulating the same gene when considering their regulatory effects. Even in this case, we suggest that the combined activity profile of those transcription factors, which can be mapped into a mathematically equivalent single transcription factor's profile, should follow our aforementioned condition when the target gene displays a cuspidate waveform.

Generally, it is known that dynamical systems with activating interactions alone do not easily generate oscillations; inhibitory interactions are also necessary. Specifically, an odd number of inhibitions need to be arranged along a feedback loop, if the loop is not too long [40-42]. In addition to this basal level of inhibitory interactions required, an abundance of cuspidate-waveform genes in the plant oscillator tips the balance in favor of a greater number of inhibitory interactions, resulting in their dominance, according to our hypothesis [cuspidate protein profiles include LHY, PRR5, TOC1, EARLY FLOWERING 4 (ELF4), LUX ARRHYTHMO (LUX), and GIGANTEA (GI) profiles in S1–S5 Fig, and comprise at least half of the available protein profiles. Among the corresponding genes, light-responsive genes are only *LHY* and *GI* (Fig 1A), which yet maintain cuspidate expression patterns in LL and DD (S4 and S5 Fig). It indicates that these patterns are largely independent of light stimulation]. For example, in loop I of the kernel, we note that both morning (*LHY/CCA1*) and evening (*TOC1*) genes show cuspidate profiles with a large phase difference between them, and are thus likely to require their own inhibitors. The simplest solution would be to have the two genes repressed by each other, but this solution, with an even number of inhibitions, would not generate oscillations. Hence, one more inhibitor, PRR5, is necessary and the subsequent introduction of the double negative connection from TOC1 to *LHY/CCA1* through PRR5, combined with the *TOC1* inhibition by LHY/CCA1, completes the repressilator structure. In addition, PRR5 should maintain a large phase difference with LHY/CCA1, because of the LHY/CCA1's cuspidate profile. Consequently, PRR5 should show a small phase difference with TOC1. Because of this small phase difference, the inhibition of *PRR5* by TOC1 cannot alone produce the empirically-observed cuspidate PRR5 profile. Therefore, *PRR5* requires an additional inhibitor with a large phase difference, LHY/CCA1. The resulting inhibition of *PRR5* by LHY/CCA1 now completes the full loop-I circuit. Through this analysis of loop I, the underlying mechanism of oscillatory dynamics with cuspidate waveforms was found to explain not only the prevalence of inhibitory interactions, but also the very specific, fine-resolution structure of loop I, revealing the loop's organizing principle.

Motivated by the intriguing connection between the shape of the waveforms and inhibitory regulation in plants, we asked if such relationships are observed in other circadian systems. Notably, a prevalence of inhibitory interactions *per se* is not conserved in other organisms: the core circadian systems of other organisms are usually simpler than those of plants, and involve feedback loops with comparable numbers of positive and negative interactions [11-13]. Those interactions are not necessarily transcriptional, and thus, caution should be taken when they are analyzed in our waveform-shape framework, which has been derived from the mathematical models of transcriptional regulation. Despite this caveat, in a preliminary analysis below, we applied our framework to both transcriptional and non-transcriptional interactions, considering their possible mathematical similarity at the coarse-grained level.

In the core circadian clock of the fungus *Neurospora crassa*, WHITE COLLAR-1, 2 (WC-1 and WC-2) proteins form a WHITE COLLAR COMPLEX (WCC) that activates the expression of *frequency* (*frq*) gene. The expressed FRQ protein subsequently blocks the WCC activity by the clearance of WC-1 [12]. In this negative feedback loop, WC-1 is suppressed by FRQ, which is upregulated by WC-1. From the experimental data [43], we observed that WC-1 exhibits a cuspidate profile, while having a large phase difference (~11 hours in DD) with FRQ. At the same time, FRQ shows a smooth sinusoidal profile. Despite multiple complicating factors in a rigorous analysis of species other than plants, this preliminary result from the *Neurospora* data is supportive of a relation between waveform-shape, phase differences, and interaction types (activation or inhibition), which is suggested by our waveform-specifying framework.

## Discussion

In this study, we explored the underlying mechanism of the plant circadian system through a systematic *in silico* analysis of the clock gene circuitry, revealing its kernel architecture to be an interaction between four negative feedback loops dominated by inhibitory regulations (Fig 1A and Fig 2). The kernel encompasses about half of the currently known interactions in the system, and they must be present to generate molecular rhythms close to WT. The other interactions not belonging to the kernel may play a role to improve the system's robustness to diverse disturbances (S1 Text), or may be required to form WT rhythms but under light conditions that have not been considered here due to limited data availability. A follow-up analysis is warranted for a more holistic understanding of plant circadian dynamics. Overall, our study illustrates the remarkable utility of mechanistic simulations, which can complement experimental approaches, in deciphering important biological processes [44-46] such as circadian rhythms.

We suggested that a preponderance of inhibitory interactions at the core of the plant clock reflects abundant cuspidate profiles of clock genes, and facilitates the global coordination of temporally-distant clock events which are sharply peaked at very specific times. We envisage that this type of cuspidate waveforms helps confer high-resolution timing to many subsequent downstream tasks in plant physiology and development [35, 47]. Whether a certain class of waveforms other than cuspidate shapes will also benefit from inhibitory interactions will be an interesting issue to address.

Besides the effect on waveforms, alternative hypotheses might be possible to explain the prevalence of the inhibitory interactions, e.g., in the context of stochasticity in molecular events, or the system's response time [48-50]. Yet, we are not aware of any explicit link or evidence to connect those mechanisms to dominant inhibitory interactions in the plant clock. Nevertheless, the possible relevance of those mechanisms deserves active investigation, towards a comprehensive picture of the plant circadian system viewed from various angles.

The four negative feedback loops within the kernel present an array of interesting predictions, which are experimentally testable. The *Δprr5/prr7* double mutation severely impairs the free running rhythmicity of clock-controlled gene expression [36]. According to our prior discussion of the loops-I-to-III inactivation, only the removal of both PRR5 and PRR7's inhibitory actions on *LHY* and *CCA1*, rather than entire deletions of *PRR5* and *PRR7*, should suffice to phenocopy the double mutant, or at least, to considerably alter clock gene expression patterns. Additionally, from the reciprocal inhibitions within the unique loop-I structure, we suggested that an increase of the *PRR5* inhibition by LHY/CCA1 would lengthen the free running period and that the opposite perturbation would shorten the period (S1 Text). Furthermore, in the context of inhibitory interactions and cuspidate waveforms, we proposed that decreasing the *PRR5* inhibition by LHY/CCA1 under 12L:12D cycles, balanced by strengthening the *PRR5* inhibition by TOC1, would reduce the peak-to-trough change in the PRR5 expression profile (S1 Text). Experimental validation of all these predictions would require manipulation of specific interactions between genes, rather than the alteration or deletion of the functionality of the entire gene itself. This could be achieved, for example, by modifying key *cis*-regulatory elements at the relevant promoter sites. Any discrepancy between experimental and computational results might be useful for our model improvement. Further consideration of protein segregation into different cellular compartments [27, 28], stochastic fluctuation in mRNA and protein concentrations [49, 51, 52], stimulus by temperature changes and endogenous sugar supply [53, 54], and tissue-specific clock regulation [55] offers additional avenues towards more complete mathematical models. Various methods to infer biological networks would also contribute to this direction [56-59]. Finally, our systematic approach advances the goal for a fundamental design principle of biological clockwork [53, 60-62], as well as for an optimal circuitry design in synthetic biology [32, 63, 64].

## Materials and Methods

**Mathematical modeling of the plant circadian system**

We constructed our mathematical model (MF2015) of the core circadian clock in *Arabidopsis* by applying system identification techniques [65]. Transcriptional, post-translational, and light regulations of molecular components were considered for model construction, primarily based on experimentally verified knowledge. The model consists of 24 ODEs employing Michaelis-Menten kinetics. Each ODE describes the concentration rate change of the corresponding mRNA, protein, or protein complex: typically, for mRNAs, $\dot{c}^m(t) = f_1[\{c^{TF}(t)\}, \{h\}, \{\theta\}] - g_1[c^m(t), \{\theta\}]$, and for proteins, $\dot{c}^p(t) = f_2[c^m(t), \{\theta\}] - g_2[c^p(t), \{\theta\}]$. Here, $c^m$ ($c^p$) denotes mRNA (protein) concentration, $c^{TF}$ denotes the transcription factor concentration, $t$ is time, the function $f_1$ ($f_2$) describes transcriptional (translational) mechanisms, the function $g_1$ ($g_2$) describes mRNA (protein) degradation, $\theta$'s are model parameters, $h$'s are the Hill coefficients, and $\{\cdots\}$ includes single or multiple elements. If experimental evidence indicates that transcription factors form a dimer, we set the Hill coefficient to be 2, otherwise, it is set to 1 [33, 66]. Transcriptional regulation in $f_1$ is modeled by $\theta_1(c^{TF})^h/[\theta_2^h+(c^{TF})^h]$ for activation or $\theta_1/[\theta_2^h+(c^{TF})^h]$ for inhibition. The regulatory effect of multiple activators (inhibitors) is modeled by the summation (product) of individual regulatory effects, with some exceptions such as PRR proteins (S1 Text) [29, 67]. We model the binding of ZTL and GI proteins by adapting the alternative Michaelis-Menten relation in [68]. For the model parameter estimation, we collected experimental time course data of mRNA and protein levels in WT *Arabidopsis* from publicly available sources listed in S1 Table. Because the absolute mRNA and protein levels were difficult to ascertain from their sources, we normalized the mRNA and protein levels into dimensionless values ($\leq 1$) with arbitrary scales (S1 Text). We compared the simulation results with experimental data and applied the prediction error method with a quadratic criterion [65] to estimate the parameters; minimization of a mean squared error between the simulated and experimental data gave rise to the estimated parameters. Before the minimization, the initial parameters were chosen using a linear least square method described in [26]. The minimization was performed using the MATLAB function *fminsearch*. In cases where constraints need to be imposed on the parameters to avoid over-fitting or biologically unrealistic solutions, the MATLAB function *fmincon* was used. Full details of the model construction, equations, and parameters are presented in S1 Text.

**Identification of the kernel structure**

In this study, a kernel is defined as a collection of spanning subgraphs that satisfy the following condition: each spanning subgraph contains all molecular components in the system and a minimal subset of their regulatory interactions (including light regulation), which are necessary to generate the temporal trajectory of molecular concentrations close to those of WT. Identification of the exact kernel demands very extensive computational resources; therefore, we used a heuristic approach to estimate the kernel structure. In this procedure, both molecular interactions and light regulations are referred to simply as interactions. First, we simulated the knockout effect of each interaction on WT expression patterns under five different light conditions. The knockout effect was quantified for each molecular component and light condition, by a root mean square error (RMSE) between the simulated mutant and WT expression profiles of the component in that light condition (S1 Text). After deletion of a given interaction, we identified the largest value ($RMSE_{max}$) among RMSEs for all components and light conditions except for GI and ZTL proteins in LL ($RMSE_{GI,LL}$ and $RMSE_{ZTL,LL}$). Based on our manual inspection, the model outputs appear to remain robust if they simultaneously satisfy $RMSE_{max} \leq 0.2$, $RMSE_{GI,LL} \leq 0.5$, and $RMSE_{ZTL,LL} \leq 0.5$ (because GI and ZTL levels are substantially elevated in LL, they allow relatively large RMSEs). From MF2015, we pruned all interactions with small knockout effects ($RMSE_{max} \leq 0.2$, $RMSE_{GI,LL} \leq 0.5$, and $RMSE_{ZTL,LL} \leq 0.5$). The simulated profiles with the only remaining interactions after the pruning still showed $RMSE_{max} \leq 0.2$, $RMSE_{GI,LL} \leq 0.5$, and $RMSE_{ZTL,LL} \leq 0.5$. Among these remaining interactions, we focused on the interactions that satisfy $RMSE_{max} \leq 0.3$, $RMSE_{GI,LL} \leq 0.8$, and $RMSE_{ZTL,LL} \leq 0.8$. We found that some of these interactions can be additionally removed from the system because the simultaneous deletion of those interactions eventually resulted in $RMSE_{max} \leq 0.2$, $RMSE_{GI,LL} \leq 0.5$, and $RMSE_{ZTL,LL} \leq 0.5$, when parameter re-optimization was performed (S1 Text). We did not attempt to delete interactions with larger $RMSE_{max}$, $RMSE_{GI,LL}$, or $RMSE_{ZTL,LL}$ ($RMSE_{max} > 0.3$, $RMSE_{GI,LL} > 0.8$, or $RMSE_{ZTL,LL} > 0.8$) with the original parameters, because these RMSEs were not usually reduced to $RMSE_{max} \leq 0.2$, $RMSE_{GI,LL} \leq 0.5$, and $RMSE_{ZTL,LL} \leq 0.5$ after parameter re-optimization. The exception to these procedures was the *PRR7* inhibition by TOC1. This interaction was removed initially because of small RMSEs caused by the deletion. In fact, the small RMSEs resulted from the *PRR7* inhibition by the EC, which buffered the loss of the inhibition by TOC1. The *PRR7* inhibition by TOC1 and that by the EC are almost equivalent to each other, because of the same target gene (*PRR7*) and regulation type (inhibition), and similar TOC1 and EC profiles in MF2015. Indeed, the removal of the *PRR7* inhibition by the EC from MF2015 was compensated for by the inhibition by TOC1 when accompanied by parameter re-optimization. Because of the equivalence of these two inhibitory interactions, we reinstated the *PRR7* inhibition by TOC1 in the kernel structure. No other interaction was reinstated due to a lack of such equivalence. The simulation of the resulting kernel structure with re-optimized parameters produces the WT expression profiles that capture the overall experimental and MF2015-simulated profiles (S1–S5 Fig). Further details of the kernel identification are presented in S1 Text.

**Modeling the relationship between cuspidate profiles and regulatory interactions**

To investigate how transcriptional regulation affects the formation of cuspidate profiles, we considered a mathematical system containing a single transcription factor (either an inhibitor or activator) and its own target gene (Fig 4A). The ODEs for this system are given by $\dot{x}_m(t) = g[x_{TF}(t), h, \{\alpha\}] - \lambda_m x_m(t)$ and $\dot{x}_p(t) = x_m(t) - \lambda_p x_p(t)$, where $x_{TF}$ denotes the transcription factor concentration, $x_m$ ($x_p$) denotes the target gene's mRNA (protein) concentration, $t$ is time, $g = x_{TF}^h/(\alpha_1 + \alpha_2 x_{TF}^h) + \alpha_3$ if the transcription factor is an activator, $g = 1/(\alpha_1 + \alpha_2 x_{TF}^h)$ if the transcription factor is an inhibitor, $h$ is the Hill coefficient, and $\alpha$'s and $\lambda$'s are constants. In the equation for $\dot{x}_p(t)$, without loss of generality, we omitted the coefficient for a protein synthesis rate per mRNA in front of $x_m(t)$. Therefore, technically, $x_m(t)$ should be interpreted as the protein synthesis rate, rather than as the mRNA concentration itself. Although the equation for $\dot{x}_m(t)$ was formulated for the case of a single transcription factor, it generally works for multiple transcription factors as well, because the combined activity profile of these transcription factors (represented by $g$) can be mapped into a mathematically equivalent single transcription factor's profile. To generate $x_p(t)$ having a cuspidate waveform schematized in Fig 4B, we considered various forms of $x_{TF}(t)$ and activating and inhibitory regulations. Given the form of $x_{TF}(t)$, we computed $x_p(t)$ with the parameters that best fit $x_p(t)$ into a cuspidate profile in Fig 4B. The resulting $x_p(t)$ was compared to Fig 4B, and their similarity was evaluated. Further details are presented in S1 Text.

## Acknowledgments

We thank Seung-Hee Yoo and Andrey Dovzhenok for useful discussions and Flora Aik for assistance with the figure preparation. This work was supported by the National Research Foundation of Korea Grant 2015R1C1A1A02037045 funded by the Korean Government (Ministry of Science, ICT and Future Planning) (to PJK), and by National Institutes of Health Grant R01GM093285 (to DES).

## References

1. Nagel DH, Kay SA. Complexity in the wiring and regulation of plant circadian networks. Curr Biol. 2012; 22:R648-R657. doi: 10.1016/j.cub.2012.07.025 PMID: 22917516
2. Sehgal A, Price J, Man B, Young M. Loss of circadian behavioral rhythms and *per* RNA oscillations in the *Drosophila* mutant *timeless*. Science. 1994; 263:1603-1606. doi: 10.1126/science.8128246
3. Brody S, Harris S. Circadian rhythms in *Neurospora*: spatial differences in pyridine nucleotide levels. Science. 1973; 180:498-500. doi: 10.1126/science.180.4085.498
4. McClung CR. Plant circadian rhythms. Plant Cell. 2006; 18:792-803. doi: 10.1105/tpc.106.040980 PMID: 16595397
5. Serkh K, Forger DB. Optimal schedules of light exposure for rapidly correcting circadian misalignment. PLoS Comput Biol. 2014; 10:e1003523. doi: 10.1371/journal.pcbi.1003523 PMID: 24722195
6. Karatsoreos IN, Bhagat S, Bloss EB, Morrison JH, McEwen BS. Disruption of circadian clocks has ramifications for metabolism, brain and behavior. Proc Natl Acad Sci USA. 2011; 108:1657-1662. doi: 10.1073/pnas.1018375108 PMID: 21220317
7. Relogio A, Westermark PO, Wallach T, Schellenberg K, Kramer A, Herzel H. Tuning the mammalian circadian clock: robust synergy of two loops. PLoS Comput Biol. 2011; 7:e1002309. doi:10.1371/journal.pcbi.1002309 PMID: 22194677
8. Ruger M, Scheer FAJL. Effects of circadian disruption on the cardiometabolic system. Rev Endocr Meta Disord. 2009; 10:245-260. doi: 10.1007/s11154-009-9122-8 PMID: 19784781
9. Harmer SL. The circadian system in higher plants. Annu Rev Plant Biol. 2009; 60:357-377. doi:10.1146/annurev.arplant.043008.092054

10. Romanovski A, Yanovsky MJ. Circadian rhythms and post-transcriptional regulation in higher plants. Front Plant Sci. 2015; 6:437. doi: 10.3389/fpls.2015.00437 PMID: 26124767

11. Robinson I, Reddy A. Molecular mechanisms of the circadian clockworks in mammals. FEBS Lett. 2014; 558:2477-2483. doi: 10.1016/j.febslet.2014.06.005

12. Dunlap JC, Loros JJ. How fungi keep time: circadian system in *Neurospora* and other fungi. Curr Opin Microbiol. 2006; 9:579-587. doi:10.1016/j.mib.2006.10.008

13. Hardin PE. Molecular genetic analysis of circadian timekeeping *Drosophila*. Adv Genet. 2011; 74:141-173. doi: 10.1016/B978-0-12-387690-4.00005-2 PMID: 21924977

14. Dunn S-J, Martello G, Yordanov B, Emmott S, Smith AG. Defining an essential transcription factor program for naïve pluripotency. Science. 2014; 344:1156-1160. doi: 10.1126/science.1248882 PMID: 24904165

15. Yang L, Tan J, O'Brien EJ, Monk JM, Kim D, Li HJ, et al. Systems biology definition of the core proteome of metabolism and expression is consistent with high-throughput data. Proc Natl Acad Sci USA. 2015; 112:10810-10815. doi: 10.1073/pnas.1501384112 PMID: 26261351

16. Papin JA, Stelling J, Price ND, Klamt S, Schuster S, Palsson BO. Comparison of network-based pathway analysis methods. Trends Biotechnol. 2004; 22:400-405. doi: 10.1016/j.tibtech.2004.06.010

17. Kim J, Park S-M, Cho K-H. Discovery of a kernel for controlling biomolecular regulatory networks. Sci Rep. 2013; 3:2223. doi: 10.1038/srep02223

18. Trinh CT, Unrean P, Srienc F. Minimal *Escherichia coli* cell for the most efficient production of ethanol from hexoses and pentoses. Appl Environ Microbiol. 2008; 74:3634-3643. doi: 10.1128/AEM.02708-07 PMID: 18424547

19. Leloup J-C, Goldbeter A. Toward a detailed computational model for the mammalian circadian clock. Proc Natl Acad Sci USA. 2003; 100:7051-7056. doi: 10.1073/pnas.1132112100 PMID: 12775757

20. Ueda HR, Hagiwara M, Kitano H. Robust oscillations within the interlocked feedback model of *Drosophila* circadian rhythm. J Theor Biol. 2001; 210:401-406. doi: 10.1006/jtbi.2000.2226

21. Hong CI, Jolma IW, Loros JJ, Dunlap JC, Ruoff P. Simulating dark expressions and interactions of *frq* and *wc-1* in the *Neurospora* circadian clock. Biophys J. 2008; 94:1221-1232. doi: 10.1529/biophysj.107.115154 PMID: 17965132

22. Locke JCW, Millar AJ, Turner MS. Modelling genetic networks with noisy and varied experimental data: the circadian clock in *Arabidopsis thaliana*. J Theor Biol. 2005; 234:383-393. doi: 10.1016/j.jtbi.2004.11.038

23. Pokhilko A, Mas P, Millar A. Modelling the widespread effects of TOC1 signalling on the plant circadian clock and its outputs. BMC Syst Biol. 2013; 7:23. doi: 10.1186/1752-0509-7-23 PMID: 23506153

24. Fogelmark K, Troein C. Rethinking transcriptional activation in the *Arabidopsis c*ircadian clock. PLoS Comput Biol. 2014; 10:e1003705. doi: 10.1371/journal.pcbi.1003705 PMID: 25033214

25. Seaton DD, Smith RW, Song YH, MacGregor DR, Stewart K, Steel G, et al. Linked circadian outputs control elongation growth and flowering in response to photoperiod and temperature. Mol Syst Biol. 2015; 11:776. doi: 10.15252/msb.20145766 PMID: 25600997

26. Foo M, Somers D, Kim P-J. System identification of the *Arabidopsis* plant circadian system. J Korean Phys Soc. 2015; 66:700-712. doi: 10.3938/jkps.66.700

27. Kim Y, Han S, Yeom M, Kim H, Lim J, Cha J-Y, et al. Balanced nucleocytosolic partitioning defines a spatial network to coordinate circadian physiology in plants. Dev Cell. 2013; 26:73-85. doi: 10.1016/j.devcel.2013.06.006

28. Wang L, Fujiwara S, Somers DE. PRR5 regulates phosphorylation, nuclear import and subnuclear localization of TOC1 in the *Arabidopsis* circadian clock. EMBO J. 2010; 29:1903-1915. doi: 10.1038/emboj.2010.76 PMID: 20407420

29. Farre EM, Harmer SL, Harmon FG, Yanovsky MJ, Kay SA. Overlapping and distinct roles of *PRR7* and *PRR9* in the *Arabidopsis c*ircadian clock. Curr Biol. 2005; 15:47-54. doi: 10.1016/j.cub.2004.12.067

30. Somers DE, Kim W-Y, Geng R. The F-Box protein ZEITLUPE confers dosage-dependent control on the circadian clock, photomorphogenesis, and flowering time. Plant Cell. 2004; 16:769-782. doi: 10.1105/tpc.016808 PMID: 14973171

31. Kim W-Y, Geng R, Somers DE. Circadian phase-specific degradation of the F-box protein ZTL is mediated by the proteasome. Proc Natl Acad Sci USA. 2003; 100:4933-4938. doi: 10.1073/pnas.0736949100 PMID: 12665620

32. Elowitz MB, Leibler S. A synthetic oscillatory network of transcriptional regulators. Nature. 2000; 403:335-338. doi: 10.1038/35002125

33. Pokhilko A, Fernández AP, Edwards KD, Southern MM, Halliday KJ, Millar AJ. The clock gene circuit in *Arabidopsis* includes a repressilator with additional feedback loops. Mol Syst Biol. 2012; 8:574. doi: 10.1038/msb.2012.6 PMID: 22395476

34. Mas P, Yanovsky MJ. Time for circadian rhythms: plants get synchronized. Curr Opin Plant Biol. 2009; 12:574-579. doi: 10.1016/j.pbi.2009.07.010

35. Huang W, Pérez-García P, Pokhilko A, Millar AJ, Antoshechkin I, Riechmann JL, et al. Mapping the core of the *Arabidopsis* circadian clock defines the network structure of the oscillator. Science. 2012; 336:75-79. doi: 10.1126/science.1219075

36. Nakamichi N, Kita M, Ito S, Sato E, Yamashino T, Mizuno T. The *Arabidopsis* Pseudo-response regulators, PRR5 and PRR7, coordinately play essential roles for circadian clock function. Plant Cell Physiol. 2005; 46:609-619. doi: 10.1093/pcp/pci061

37. Alabadi D, Yanovsky MJ, Mas P, Harmer SL, Kay SA. Critical role for CCA1 and LHY in maintaining circadian rhythmicity in *Arabidopsis*. Curr Biol. 2002; 12:757-761. doi: 10.1016/S0960-9822(02)00815-1

38. Mizoguchi T, Wheatley K, Hanzawa Y, Wright L, Mizoguchi M, Song H-R, et al. *LHY* and *CCA1* are partially redundant genes required to maintain circadian rhythms in *Arabidopsis*. Dev Cell. 2002; 2:629-641. doi: 10.1016/S1534-5807(02)00170-3

39. Somers DE. The *Arabidopsis* clock: time for an about-face? Genome Biol. 2012; 13:153. doi: 10.1186/gb-2012-13-4-153 PMID: 22546533

40. Patel VR, Eckel-Mahan K, Sassone-Corsi P, Baldi P. How pervasive are circadian oscillations? Trends Cell Biol. 2014; 24:329-331. doi: 10.1016/j.tcb.2014.04.005

41. Goh K-I, Kahng B, Cho K-H. Sustained oscillations in extended genetic oscillatory systems. Biophys J. 2008; 94:4270-4276. doi: 10.1529/biophysj.107.128017 PMID: 18326637

42. Friesen WO, Friesen JA. NeuroDynamix II: Concepts of Neurophysiology Illustrated by Computer Simulations, 2nd Edition. NY, New York: Oxford University Press; 2009.

43. Merrow M, Franchi L, Dragovic Z, Görl M, Johnson J, Brunner M, et al. Circadian regulation of the light input pathway in *Neurospora crassa*. EMBO J. 2001; 20:307-315. doi: 10.1093/emboj/20.3.307 PMID: 11157738

44. Goodfellow M, Phillips NE, Manning C, Galla T, Papalopulu N. microRNA input into a neural ultradian oscillator controls emergence and timing of alternative cell states. Nat Commun. 2014; 5:3399. doi: 10.1038/ncomms4399 PMID: 24595054

45. Mehra A, Hong CI, Shi M, Loros JJ, Dunlap JC, Ruoff P. Circadian rhythmicity by autocatalysis. PLoS Comput Biol. 2006; 2:e96. doi:10.1371/journal.pcbi.0020096 PMID: 16863394

46. Forger DB, Peskin CS. Stochastic simulation of the mammalian circadian clock. Proc Natl Acad Sci USA. 2005; 102:321-324. doi: 10.1073/pnas.1205156109 PMID: 23027938

47. Nakamichi N, Kiba T, Kamioka M, Suzuki T, Yamashino T, Higashiyama T, et al. Transcriptional repressor PRR5 directly regulates clock-output pathways. Proc Natl Acad Sci USA. 2012; 109:17123-17128. doi: 10.1073/pnas.1205156109 PMID: 23027938

48. Shinar G, Dekel E, Tlusty T, Alon U. Rules for biological regulation based on error minimization. Proc Natl Acad Sci USA. 2006; 103:3999-4004. doi: 10.1073/pnas.0506610103 PMID: 16537475

49. Rao CV, Wolf DM, Arkin AP. Control, exploitation and tolerance of intracellular noise. Nature. 2002; 420:231-237. doi: 10.1038/nature01258

50. Mangan S, Itzkovitz S, Zaslaver A, Alon U. The incoherent feed-forward loop accelerates the response-time of the *gal* system of *Escherichia coli*. J Mol Biol. 2006; 356:1073-1081. doi: 10.1016/j.jmb.2005.12.003

51. Raj A, van Oudenaarden A. Nature, nurture, or chance: stochastic gene expression and its consequences. Cell. 2008; 135:216-226. doi: 10.1016/j.cell.2008.09.050 PMID: 18957198

52. Kim P-J, Price ND. Macroscopic kinetic effect of cell-to-cell variation in biochemical reactions. Phys Rev Lett. 2010; 104:148103. doi: 10.1103/PhysRevLett.104.148103

53. Gould PD, Locke JCW, Larue C, Southern MM, Davis SJ, Hanano S, et al. The molecular basis of temperature compensation in the *Arabidopsis* circadian clock. Plant Cell. 2006; 18:1177-1187. doi: 10.1105/tpc.105.039990 PMID: 16617099

54. Haydon MJ, Mielczarek O, Robertson FC, Hubbard KE, Webb AAR. Photosynthetic entrainment of the *Arabidopsis thaliana* circadian clock. Nature. 2013; 502:689-692. doi: 10.1038/nature12603 PMID: 24153186

55. Endo M, Shimizu H, Nohales MA, Araki T, Kay SA. Tissue-specific clocks in *Arabidopsis* show asymmetric coupling. Nature. 2014; 515:419-422. doi: 10.1038/nature13919 PMID: 25363766

56. Faith JJ, Hayete B, Thaden JT, Mogno I, Wierzbowski J, Cottarel G, et al. Large-scale mapping and validation of *Escherichia coli* transcriptional regulation from a compendium of expression profiles. PLoS Biol. 2007; 5:e8. doi: 10.1371/journal.pbio.0050008 PMID: 17214507

57. Vicente R, Wibral M, Lindner M, Pipa G. Transfer entropy—a model-free measure of effective connectivity for the neurosciences. J Comput Neurosci. 2011; 30:45-67. doi: 10.1007/s10827-010-0262-3 PMID: 20706781

58. Sohl-Dickstein J, Battaglino PB, DeWeese MR. New method for parameter estimation in probabilistic models: minimum probability flow. Phys Rev Lett. 2011; 107:220601. doi: 10.1103/PhysRevLett.107.220601

59. Kim P-J, Price ND. Genetic co-occurrence network across sequenced microbes. PLoS Comput Biol. 2011; 7:e1002340. doi: 10.1371/journal.pcbi.1002340 PMID: 22219725

60. Harmer SL, Panda S, Kay SA. Molecular bases of circadian rhythms. Annu Rev Cell Dev Biol. 2001; 17:215-253. doi: 10.1146/annurev.cellbio.17.1.215

61. Millar AJ. Input signals to the plant circadian clock. J Exp Bot. 2004; 55:277-283. doi: 10.1093/jxb/erh034

62. Webb A. The physiology of circadian rhythms in plants. New Phytol. 2003; 160:281-303. doi: 10.1046/j.1469-8137.2003.00895.x

63. Andrianantoandro E, Basu S, Karig DK, Weiss R. Synthetic biology: new engineering rules for an emerging discipline. Mol Syst Biol. 2006; 2:2006.0028. doi: 10.1038/msb4100073 PMID: 16738572

64. Khalil AS, Collins JJ. Synthetic biology: applications come of age. Nat Rev Genet. 2010; 11:367-379. doi: 10.1038/nrg2775 PMID: 20395970

65. Ljung L. System Identification: Theory for the User, 2nd Edition. Englewood Cliff, NJ: Prentice Hall; 1999.

66. Heck HdA. Statistical theory of cooperative binding to proteins. Hill equation and the binding potential. J Am Chem Soc. 1971; 93:23-29. doi: 10.1021/ja00730a004

67. Nakamichi N, Kiba T, Henriques R, Mizuno T, Chua N-H, Sakakibara H. PSEUDO-RESPONSE REGULATORS 9, 7, and 5 are transcriptional repressors in the *Arabidopsis* circadian clock. Plant Cell. 2010; 22:594-605. doi: 10.1105/tpc.109.072892 PMID: 20233950

68. Levine E, Hwa T. Stochastic fluctuations in metabolic pathways. Proc Natl Acad Sci USA. 2007; 104:9224-9229. doi: 10.1073/pnas.0610987104 PMID: 17517669